\documentclass[preprint,12pt,authoryear]{elsarticle}

\usepackage{amssymb}
\usepackage{amsmath}
\usepackage{color} 
\usepackage{comment} 
\usepackage{soul} 

\usepackage{lineno}

\journal{Ocean Modelling}

\begin{document}

\begin{frontmatter}



\title{A Simple and Transparent Method for Improving the Energetics and Thermodynamics of Seawater Approximations: Static Energy Asymptotics (SEA)}


\author[inst1]{R\'emi Tailleux}

\affiliation[inst1]{organization={Department of Meteorology, University of Reading},
            addressline={Whiteknights road, Earley Gate}, 
            city={Reading},
            postcode={RG6 6ET}, 
            state={Berkshire},
            country={United Kingdom}}

\author[inst2]{Thomas Dubos}

\affiliation[inst2]{organization={IPSL, Lab. Meteorologie Dynamique, 
Ecole Polytechnique},
            city={Palaiseau},
            country={France}}

\begin{abstract}
The static energy encodes all possible information about the thermodynamics and potential energy (and all related forces) of stratified geophysical fluids. In this paper, we develop a systematic methodology, called static energy asymptotics, that exploits this property for constructing energetically and thermodynamically consistent sound-proof approximations of the equations of motion. By approximating the static energy to various orders of accuracy, two main families of approximations are (re-)derived and discussed: the pseudo-incompressible (PI) approximation and the anelastic (AN) approximation. \textcolor{black}{For all approximations, the background and available potential energies (in Lorenz sense) can be constructed to match their exact counterparts as closely as feasible and to be expressible in terms of the exact (as opposed to ad-hoc) thermodynamic potentials.} For hydrostatic motions, the AN approximation (of which the Boussinesq approximation is a special case) has the same structure as that of legacy Seawater Boussinesq primitive equations. The energetics of such models could therefore be made transparently traceable to that of the full Navier-Stokes equations at little to no additional cost, thus allowing them to take
full advantage of the Gibbs Sea Water (GSW) library 
\textcolor{black}{developed as part of the new thermodynamic standard for seawater TEOS-10}. 
\end{abstract}


\begin{highlights}
\item New energetically and thermodynamically consistent approximations are constructed
\item New approach is based on approximating static energy to various orders of accuracy
\item Approximate forms of potential energy are transparently related to exact counterpart
\item Legacy standard seawater Boussinesq primitive equations are needlessly inaccurate 
\item Existing Boussinesq ocean models could easily be improved at little additional cost
\end{highlights}


\end{frontmatter}



\section{Introduction} 

Since their inception in the late 60's, numerical ocean models have almost exclusively relied on the so-called Seawater Boussinesq approximation (SBA thereafter). Unlike the standard Boussinesq approximation, the SBA makes use of the full nonlinear equation of state for seawater and therefore retains the adiabatic and isohaline compressibility effects resulting from its pressure dependence. Doing so is essential because these nonlinearities appear to be crucial for correctly simulating the relative layering of ocean water masses, e.g., \citet{Nycander2015,IOC2010}. The nonlinearities most important dynamically are: the {\em cabbeling nonlinearity}, which is associated with the temperature dependence of the thermal expansion coefficient and responsible for the densification upon mixing, and the {\em thermobaric nonlinearity}, which is associated with the pressure dependence of the thermal expansion coefficient and responsible for making colder parcels more compressible than warmer ones \citep{McDougall1987c}.

Although the SBA has received much attention, its accuracy, exact justification, as well as its energetic and thermodynamic consistency, have remained poorly understood and sources of confusion. Conceptually, the construction and justification of the SBA used in numerical ocean models a priori involves a two-step process: 1) the first one pertaining to the coarse-graining of the un-averaged equations of motion; 2) the second one pertaining to the sound-proofing of the equations. So far, most approaches have generally assumed that 2) should be carried out before 1), as it is easiest to implement and understand, but the opposite view also exists. For instance, \citet{McDougall2002} have argued that the divergence-free velocity field used by the SBA should be interpreted as the Favre averaged velocity field $\overline{\bf v}^{\rho} = \overline{\rho {\bf v}}/\overline{\rho}$, thus regarding the SBA as being primarily the result of the coarse-graining procedure rather than of the Boussinesq sound-proofing step. Clarifying the issue is important for establishing the accuracy of the SBA, because in McDougall's interpretation, the continuity equation $\nabla \cdot \overline{\bf v}^{\rho} = 0$ is exact in a steady-state, whereas it is only approximate in the conventional interpretation. 

In this paper, our main aim is to clarify the key ingredients (and those that are not) of any construction of energetically and thermodynamically consistent approximations, defined here as approximations that admit an energy conservation principle and thermodynamic potentials traceable in a rigorous and physically transparent way to their exact counterparts, and whose description of irreversible processes/diabatic effects remains in agreement with the second law of thermodynamics. To that end, we address the issue in the context of the un-averaged compressible Navier-Stokes equations, as how coarse-graining affects energetics and thermodynamics is currently much less understood. Note that in the literature, the concept of energetically and thermodynamically consistent models is also occasionally understood as models predicting the evolution of coarse-grained motions whose turbulent closures are consistent with energy conservation and the second law of thermodynamics. Those models may also include prognostic equations for subgridscale energy reservoirs, e.g., \citet{Eden2014,Eden2015,Eden2016}. 

Following relatively recent progress \citep{Young2010,Tailleux2010c,Tailleux2012,Eden2014,Eden2015,Eden2016}, the SBA is now understood to admit a well defined energy conservation principle, although one that requires the construction of ad-hoc thermodynamic potentials. One of the main results of this paper is that this undesirable feature stems from the un-necessary character of some of the approximations/simplifications made by the SBA. Indeed, we find that it is always possible to construct SBA-like approximations whose thermodynamics can be expressed in terms of the exact thermodynamic potentials. The approximations that we propose, unlike the legacy SBA, can therefore take full advantage of the Gibbs Seawater Library (GSW) developed as part of the new international thermodynamic standard TEOS-10 \citep{IOC2010}. 

It has long been observed that the SBA can be regarded as being isomorphic to the fully compressible hydrostatic equations of motion written in pressure coordinates \citep{deSzoeke2002}. This result is important, because it implies that SBA-model can in principle be made fully compressible and therefore more accurate simply by re-interpreting height as pressure and implementing a few other changes without fundamentally altering the architecture of existing codes, as discussed and tested by \citet{Losch2004} for instance. The use of pressure coordinates is much less natural for the oceans than it is for the atmosphere, however, since it requires considering a prognostic equation for the bottom pressure (which is difficult to observe) rather than for the free surface (observable by satellite altimetry). For this reason, there is a strong incentive to continue searching for alternative ways to improve on the SBA that retain standard coordinates. To that end, there appears to be a wealth of sound-proof approximations to choose from. The anelastic system \citep{Ogura1962, lipps_scale_1982} has been initially derived for dry air modelled as an ideal perfect gas and adiabatic conditions. \citet{scinocca_nonlinear_1992} have obtained a Hamiltonian formulation of the anelastic system, which has later been extended to accomodate arbitrary multicomponent thermodynamics \citep{Pauluis2008}. \citet{Vasil2013, tort_usual_2014, cotter_variational_2014} have developed variational formulations for these systems that enable to trace back their conservation properties to symmetries of an adequate Lagrangian. \citet{durran_improving_1989, Durran2008} has proposed a different approximation, known as the pseudo-incompressible (PI) approximation in an atmospheric context. Arbitrary thermodynamics and diabatic effects (i.e. molecular conduction and diffusion) have been consistently introduced in this system \citep{Klein2012}. In an oceanic context, \citet{Dewar2016} have proposed two types of ``semi-compressible'' approximations with a consistent treatment of diabatic effects, the type-I approximation coinciding with the PI system and type-II with the anelastic system. 


Most recently, \citet{Eldred2021} have devised variational \textcolor{black}{formulations} of the PI and anelastic systems that extend \citet{Vasil2013} and \citet{tort_usual_2014} by including diabatic effects that previously had to be dealt with directly at the level of the equations of motion \citep{Pauluis2008, Klein2012, Dewar2016}. Based on the existing literature, the problem of how to construct energetically and thermodynamically consistent sound-proof approximations to the equations of motion appears to be largely understood, at least in the context of the un-averaged equations of motion. In most cases, however, such approaches require a significant amount of mathematical background, e.g. differential geometry and variational principles with non-holonomic constraints. In the context of the SBA, the methods developed by \citet{Tailleux2012}, \citet{Eden2014,Eden2015,Eden2016} to achieve energetics and thermodynamic consistency are arguably much simpler, but unfortunately not sufficiently general to be easily extended to the pseudo-incompressible or anelastic approximations. As a result, there is still a strong incentive to look for alternative approaches that can make the process of constructing consistent approximations more physically intuitive while keeping the maths as elementary as feasible. The main aim of this paper is to reformulate the mathematical apparatus of \citet{Eldred2021} in a much simpler unifying formalism, where the main vehicle of the various levels of approximations is the so-called static energy. This concept, which should be more familiar to both oceanographers and atmospheric scientists, is found to allow for a transparent discussion of both the energetics and accuracy of Boussinesq-like approximations.
The hope is that by demystifying energetics and thermodynamic consistency issues it might facilitate the adoption of more straightforwardly consistent and/or more accurate approximations by numerical ocean modellers and others.

The paper is organised as follows: Section \ref{where_is_the_info} introduces the topic by first 
reminding the reader of where to locate all the information about the thermodynamics and energetics
in the compressible Navier-Stokes equations. This leads us to identify the static energy as the
most natural variable in which to encode all thermodynamic information about the fluid, and therefore
the one to approximate. Section \ref{pseudo_incompressible} shows how to construct the 
pseudo-incompressible approximation. Section \ref{generalised_anelastic} discusses the 
construction of a modernised anelastic approximation. Section \ref{modernised_sba} discusses the construction of modernised versions of the Boussinesq and anelastic approximations for seawater within a single unifying framework. 
Section \ref{ape_sba} discusses the impact of the approximation on the partitioning of the potential energy into \textcolor{black}{available potential energy (APE) and background potential energy (BPE)}, as per \citet{Tailleux2018} local formulation of \citet{Lorenz1955} 
theory of available potential energy. Section \ref{summary_and_discussion} summarises and discusses the results. 

\section{Energy conservation and structure of static energy} 

\label{where_is_the_info}

To understand how to construct energetically and thermodynamically consistent \textcolor{black}{approximations}, it is first important to understand how the equations of motion encode the information about thermodynamics and potential energy. It is also important to understand how this information might get partially lost or scrambled in the simplification or approximation process underlying the construction of idealised models or of sound-proof approximations, or even through the use of non-canonical variables. Indeed, this is necessary for identifying possible information recovery strategies. The following sections discuss and illustrate these issues.  

\label{seawater_boussinesq_approximation} 

\subsection{Remark on energy conservation in the Navier-Stokes equations}

We take as our ground-truth and reference model the Navier-Stokes equations for compressible seawater approximated as a two-constituent fluid, which may be written
\begin{equation}
     \frac{D{\bf v}}{Dt} + 2 {\bf \Omega} \times {\bf v}
 + \frac{1}{\rho} \nabla p = - \nabla \Phi + {\bf F} ,
 \label{momentum_balance} 
\end{equation}
\begin{equation}
     \frac{\partial \rho}{\partial t} + \nabla \cdot ( 
     \rho {\bf v} ) = 0  \qquad {\rm or\,\,\, equivalently} 
     \qquad \nabla \cdot {\bf v} = \frac{D}{Dt} \ln{\nu},
     \label{continuity}
\end{equation}
\begin{equation}
     \frac{D\eta}{Dt} = \dot{\eta} = 
     -\frac{1}{\rho} \nabla \cdot (\rho {\bf J}_{\eta} ) 
     + \dot{\eta}_{irr} 
     \label{entropy_budget} 
\end{equation}
\begin{equation}
      \frac{DS}{Dt} = \dot{S} = -\frac{1}{\rho} \nabla \cdot ( \rho {\bf J}_S) ,
      \label{salinity_budget} 
\end{equation}
\begin{equation}
    T = T(\eta,S,p) = \frac{\partial h}{\partial \eta}, \qquad \mu = \mu(\eta,S,p) = \frac{\partial h}{\partial S} 
    \label{thermo_variables} 
\end{equation}
\begin{equation}
     \nu = \nu(\eta,S,p) = \frac{\partial h}{\partial p} 
     \label{eos} 
\end{equation}
Here, $h=h(\eta,S,p)$ is the specific enthalpy, 
${\bf v}=(u,v,w)$ is the three-dimensional velocity field, $\rho$ is density, \textcolor{black}{$\nu = 1/\rho$ is the specific volume}, $p$ is pressure, $\Phi = g z$ is the geopotential, \textcolor{black}{${\bf \Omega}$ is Earth rotation vector}, $g$ is the acceleration of gravity, $z$ is height, $S$ is salinity, $\eta$ is specific entropy, with ${\bf J}_S$  and ${\bf J}_{\eta}$ the molecular fluxes of salt and entropy, respectively. The term \textcolor{black}{$\dot{\eta}_{irr}\ge 0$}
represents the irreversible rate of \textcolor{black}{specific} entropy production, which the second law of thermodynamics requires to be \textcolor{black}{non-negative}. 
Finally, ${\bf F} = \rho^{-1} \nabla \cdot {\bf S}$ is the viscous force, 
\textcolor{black}{where ${\bf S}$ is the deviatoric stress tensor}.

Consistency with the second law of thermodynamics is associated with the evolution equations for specific entropy and salinity (\ref{entropy_budget}) and (\ref{salinity_budget}) through the specification of the phenomenological laws for the molecular diffusive fluxes ${\bf J}_{\eta}$ and ${\bf J}_S$, as well as with the constraint that \textcolor{black}{$\dot{\eta}_{irr}\ge 0$ be positive}. From non-equilibrium thermodynamics theory, we know that the molecular diffusive fluxes act to relax the fluid towards a resting state $({\bf v}=0)$ with homogeneous in-situ temperature \textcolor{black}{$T={\rm constant}$}
and relative chemical potential $\mu = {\rm constant}$. The simplest way to achieve this is via assuming that the molecular fluxes of $\eta$ and $S$ be linearly related to the generalised forces $\nabla T/T$ and $\nabla \mu/T$ as follows
\begin{equation}
     {\bf J}_{\eta} = - L_{\eta \eta} \frac{\nabla T}{T} - L_{\eta s} 
     \frac{\nabla \mu}{T} 
     \label{phenomenological_entropy}
\end{equation}
\begin{equation}
     {\bf J}_S = - L_{\eta s} \frac{\nabla T}{T} - L_{ss} \frac{\nabla \mu}{T} 
     \label{phenomenological_salt} 
\end{equation}
with $L_{\eta \eta}$, $L_{ss}$, $L_{\eta s}$ and $L_{s\eta}$ to be determined empirically from laboratory measurements. Energy conservation can be shown to impose that the following quantity 
\begin{equation}
     \rho [ {\bf v} \cdot {\bf F} + T \dot{\eta} + \mu \dot{S} ] = - \nabla \cdot (\rho {\bf J}_E) 
     \label{conservative_constraint_on_diabatism} 
\end{equation}
be equal to the divergence of an appropriately \textcolor{black}{defined} energy flux ${\bf J}_E$. This in turn constrains the non-viscous irreversible entropy production to take the form
\begin{equation}
       \dot{\eta}_{irr}^{\rm diff} = 
      L_{\eta \eta} \frac{|\nabla T|^2}{T^2} 
      + L_{ss} \frac{|\nabla \mu|^2}{T^2}
      + (L_{s\eta} + L_{\eta s} ) \frac{\nabla T \cdot \nabla \mu}{T^2} 
\end{equation}
which is \textcolor{black}{non-negative} provided that
\begin{equation}
    L_{\eta \eta} > 0, \qquad L_{ss} > 0, \qquad L_{\eta \eta} L_{ss} - L_{\eta s} L_{s\eta} \ge 0 .
\end{equation}
Finally, the Onsager reciprocity relationship allows for the simplification $L_{\eta s} = L_{s\eta}$. As shown in the following, the main effect of a sound-proof approximation is to modify the form of the in-situ temperature and relative chemical potential entering the phenomenological laws (\ref{phenomenological_entropy}) and (\ref{salinity_budget}) but not the laws themselves. 

As is well known, in addition to being consistent with the second law of thermodynamics, 
the Navier-Stokes equations are \textcolor{black}{built} to satisfy the law of energy conservation, the conserved total energy \textcolor{black}{per unit mass} being $E_{tot} = E_k + E_p$,
with $E_k = {\bf v}^2/2$ the kinetic energy (KE), and 
\begin{equation}
    E_p = \Phi(z) + h - \frac{p}{\rho} 
    = \Phi(z) + h - p \frac{\partial h}{\partial p} 
\end{equation}
the potential energy (PE), itself the sum of gravitational potential energy $\Phi(z) = gz$ and internal energy $e = h-p/\rho$. Note that the Navier-Stokes equations contain only information about the three partial derivatives of $h$, 
\begin{equation}
     \frac{\partial h}{\partial p} = \nu = \frac{1}{\rho}, \qquad
     \frac{\partial h}{\partial \eta} = T, \qquad
     \frac{\partial h}{\partial S} = \mu ,
\end{equation}
meaning that they are only sufficient to recover $h$ up to an irrelevant integration constant. (Note though that while $\nu$ and $T$ are unambiguously defined, $\mu$ is only defined up to an arbitrary constant; it follows that $h$ is fundamentally defined only up to a linear function of $S$, see \citet{IOC2010}). Partial loss of information about $h$ commonly occurs in many systems of equations used in practice, however. This is the case, for instance, when restricting attention to adiabatic and isohaline motions ($\dot{\eta} =\dot{S} = 0$), for which knowledge of the thermohaline derivatives of $h$ ($T$ and $\mu$) is no longer formally needed for integrating the equations forward in time. In this case, the Navier-Stokes equations reduce to the Euler equations \textcolor{black}{(assuming that viscous processes are also neglected)}, and only contain enough information to recover $h$ (or $e$) up to an indeterminate function of $(\eta,S)$. In the literature, the latter are often referred to as energy Casimirs, e.g., \citet{Shepherd1993}. 

Physically, it is important to remark that specific enthalpy acts as a thermodynamic potential encoding all possible thermodynamic information only if formulated in terms of its canonical variables $(\eta,S,p)$ \citep{Alberty1994}. It follows that thermodynamic information may also be lost when using expressions of specific enthalpy formulated in terms of non-canonical variables such as potential temperature $\theta$ or Conservative Temperature $\Theta$ in place of entropy as $h=\hat{h}(S,\theta,p)$ or $\tilde{h}(S,\Theta,p)$, as is usually preferred in oceanographic practice. To avoid losing information in that case, it becomes necessary to also supply the passage relations $\eta = \hat{\eta}(S,\theta,p)$ or $\eta = \tilde{\eta}(S,\Theta,p)$ allowing one to reconstruct $h$ in terms of its canonical variables, e.g., \citet{IOC2010}. This is needed, for instance, to compute variables such as in-situ temperature $T$, whose expression is  
\begin{equation}
    T = \frac{\partial h}{\partial \eta} = 
    \frac{\partial \hat{h}}{\partial \theta} \bigg/ \frac{\partial \hat{\eta}}{\partial \theta} ,
\end{equation} 
which clearly requires knowledge of both $\hat{h}$ and $\hat{\eta}$.
In practice, the loss of thermodynamic information affecting widely used systems of equations such as the standard Seawater Boussinesq approximation used by ocean modellers and discussed next may therefore have multiple origins that one needs to be aware of, as it is obviously a pre-requisite for identifying \textcolor{black}{information} recovery strategies.

\subsection{Motivating problem: seawater Boussinesq approximation}

To make the above ideas more concrete, it is useful to discuss the particular case of the Seawater Boussinesq approximation (SBA) 
used in a majority of numerical Ocean General Circulation Models (OGCMs) and the main focus of this paper. Its governing equations are
\begin{equation}
    \frac{D{\bf u}}{Dt} + f {\bf k} \times {\bf u} + 
    \textcolor{black}{\nabla_h}
    \left (\frac{\delta p}{\rho_b} \right )  = {\bf F}_h 
    \label{SBA_momentum} 
\end{equation}
\begin{equation}
   \frac{\partial}{\partial z} 
   \left ( \frac{\delta p}{\rho_b} \right )  = b_{bou} ,
   \label{SBA_hydrostatic}
\end{equation}
\begin{equation} 
   b_{bou} = b_{bou}(S,\theta,z) = -\frac{g(\rho - \rho_b)}{\rho_b} \label{SBA_buoyancy}
\end{equation}
\begin{equation}
   \nabla \cdot ( \rho_b {\bf v} ) = 0 
   \label{incompressible_assumption}
\end{equation}
\begin{equation}
    \frac{DS}{Dt} = \textcolor{black}{-\nabla \cdot {\bf F}_S} , 
    \qquad \frac{D\theta}{Dt} = \textcolor{black}{-\nabla \cdot {\bf F}_{\theta}} 
\end{equation}
\begin{equation}
    \rho = \rho(S,\theta,p_b(z)) 
    \label{SBA_eos} 
\end{equation}
where $\rho_b$ is the constant Boussinesq reference density, which in the NEMO OGCM is generally chosen to be $\rho_b = 1026 kg m^{-3}$, $p_b = -\rho_b g z$ is the Boussinesq pressure, $\delta p = p - p_b(z)$. As $\rho_b$ is constant, the continuity equation is of course equivalent to the usual form $\nabla \cdot {\bf v} = 0$.
\textcolor{black}{Remaining notations are: ${\bf u}=(u,v)$ is the horizontal component of the coarse-grained 3D velocity field ${\bf v}=(u,v,w)$; $\nabla_h$ is the horizontal component of the $\nabla$ operator; $f$ is the Coriolis parameter; ${\bf k}$ is the unit normal vector pointing upward; ${\bf F}_h$ is the horizontal component of the turbulent viscous force; ${\bf F}_S$ and ${\bf F}_{\theta}$ are the parameterised turbulent fluxes of salt and heat.}

Although (\ref{SBA_momentum}-\ref{SBA_eos}) has formed the basis for a majority of OGCMs, that they admit an energy conservation principle was only clarified relatively recently by \citet{Young2010}, \citet{Tailleux2012}, and \citet{Eden2015} among others, building upon previous ideas by \citet{Ingersoll2005} and \citet{Pauluis2008}. Starting from the kinetic energy equation, 
\begin{equation}
    \textcolor{black}{\rho_b}  \frac{D}{Dt} \frac{{\bf u}^2}{2} + 
    \nabla \cdot ( \delta p {\bf v} ) = \textcolor{black}{\rho_b} 
    b_{bou}(S,\theta,z) \frac{Dz}{Dt} + 
    \textcolor{black}{\rho_b} {\bf F}_h\cdot {\bf u} ,
    \label{kinetic_energy_seawater}
\end{equation}
obtained by taking the scalar product of (\ref{SBA_momentum}) with ${\bf u}$ combined with (\ref{SBA_hydrostatic}) multipled by $Dz/Dt$, the key step here is to recognise that the term $b_{bou} Dz/Dt$ is the Lagrangian derivative of a pseudo potential energy $E_{p,bou}$, so that (\ref{kinetic_energy_seawater}) may be rewritten in the form
\begin{equation}
     \rho_b \frac{D}{Dt} \left ( \frac{{\bf u}^2}{2} + E_{p,bou} \right ) 
     + \nabla \cdot (\delta p {\bf v}) = 
     \rho_b \left ( \Upsilon_S \frac{DS}{Dt} + \Upsilon_{\theta} 
    \frac{D\theta}{Dt} + 
    \textcolor{black}{{\bf F}_h \cdot {\bf u}} \right )
    \label{SBA_ECP}
\end{equation}
where the Boussinesq pseudo potential energy $E_{p,bou}$ and the thermodynamic efficiencies 
$\Upsilon_S$ and $\Upsilon_{\theta}$ are defined by 
\begin{equation}
     E_{p,bou} = -\int_{z_{\star}}^z b_{bou}(S,\theta,\tilde{z}) \,{\rm d}\tilde{z} ,
\end{equation}
\begin{equation}
     \Upsilon_S = - \int_{z_{\star}}^z \frac{\partial b_{bou}}{\partial S}
     (S,\theta,\tilde{z}) \,{\rm d}\tilde{z} ,
\end{equation}
\begin{equation}
      \Upsilon_{\theta} = -\int_{z_{\star}}^z \frac{\partial b_{bou}}{\partial \theta} 
      (S,\theta,\tilde{z}) \,{\rm d}\tilde{z} ,
\end{equation}
where $z_{\star}$ is an arbitrary reference depth. Eq. (\ref{SBA_ECP}) defines an energy conservation principle for ${\bf u}^2/2 + E_{p,bou}$ for inviscid motions in the absence of diabatic sources/sinks of $\theta$ and $S$, but is unsatisfactory because $E_{p,bou}$ represents only one chunk of the total potential energy, whose determination is rendered ambiguous by the arbitrariness of $z_{\star}$ and whose link to its exact counterpart is unclear. While progress has been made towards resolving these issues, existing ideas still remain largely ad-hoc and lacking in generality, however, thus motivating the considerations developed in subsequent sections.

\subsection{Importance of static energy} 

To ensure that the full energy conservation can be recovered from the equations of motion, we find it essential to explicitly articulate the connections between aspects of the momentum balance equations and potential energy. In this paper, this is achieved by expressing the pressure-geopotential gradient force $\rho^{-1}\nabla p+\nabla \Phi$, potential energy $E_p$, and dynamics/thermodynamics coupling in terms of the static energy 
\begin{equation}
     \Sigma_{FC} (\eta,S,p,\Phi) = h(\eta,S,p) + \Phi , 
     \label{Sigma_FC}
\end{equation}
by means of the following relations:
\begin{equation}
     \frac{1}{\rho} \nabla p + \nabla \Phi =
     \frac{\partial \Sigma_{FC}}{\partial p} \nabla p 
     + \frac{\partial \Sigma_{FC}}{\partial \Phi}\nabla \Phi 
     = {\bf F}_{dyn} , 
     \label{crocco_dynamics} 
\end{equation}
\begin{equation}
     E_p = \Phi + h - \frac{p}{\rho} = 
     \Sigma_{FC} - p \frac{\partial \Sigma_{FC}}{\partial p} ,
     \label{crocco_potential} 
\end{equation}
\begin{equation}
      \underbrace{\nabla \Sigma_{FC} - \frac{\partial \Sigma_{FC}}{\partial \eta} 
      \nabla \eta - \frac{\partial \Sigma_{FC}}{\partial S} \nabla S}_{thermodynamics}  
      =
      \underbrace{\frac{\partial \Sigma_{FC}}{\partial p} \nabla p + 
      \frac{\partial \Sigma_{FC}}{\partial \Phi} \nabla \Phi}_{dynamics} ,
      \label{crocco_theorem} 
\end{equation}
where in the case of a fully compressible fluid, the partial derivatives of $\Sigma_{FC}$ are
\begin{equation}
     \frac{\partial \Sigma_{FC}}{\partial p} =\nu = \frac{1}{\rho}, 
     \quad \frac{\partial \Sigma_{FC}}{\partial \Phi} = 1 , \quad
     \frac{\partial \Sigma_{FC}}{\partial \eta} = T, \quad
     \frac{\partial \Sigma_{FC}}{\partial S} = \mu . 
\end{equation}

\textcolor{black}{Notice that in Eq. \eqref{Sigma_FC}, $(p, \eta, S, \Phi)$ are independent coordinates in an augmented state space that extends the standard thermodynamic state space $(p,\eta,S)$. By a standard abuse of notations, we use the same notation for these independent variables and for the corresponding fields.}

In the literature, the identity (\ref{crocco_theorem}) is sometimes called the Crocco-Vazsonyi theorem \citep{Crocco1937,Vazsonyi1945} (referred as the CVT thereafter) and is key for connecting the dynamics (\ref{crocco_dynamics}) to the thermodynamics. Physically, it implies that the momentum balance equations, whose standard dynamical form is
\begin{equation}
   \frac{D{\bf v}}{Dt} + 2 {\bf \Omega} \times {\bf v} 
   + \frac{\partial \Sigma_{FC}}{\partial p} \nabla p  + 
   \frac{\partial \Sigma_{FC}}{\partial \Phi} \nabla \Phi = {\bf F}
   \label{normal_nse} 
\end{equation}
may also be written in thermodynamic form as follows: 
\begin{equation}
   \frac{D{\bf v}}{Dt} + 2 {\bf \Omega} \times {\bf v} 
   + \nabla \Sigma_{FC} = {\bf F} + \frac{\partial \Sigma_{FC}}{\partial \eta} 
   \nabla \eta + \frac{\partial \Sigma_{FC}}{\partial S} \nabla S ,
   \label{crocco_nse} 
\end{equation}
referred to as the Crocco equations thereafter. As shown throughout this paper, 
the three identities (\ref{crocco_dynamics}), (\ref{crocco_potential}), and
(\ref{crocco_theorem}) greatly facilitate understanding of energy conservation issues. 
The key idea of this paper is to argue that to understand the energetics and thermodynamic
consistency of any given approximation to the equations of motion, it is essential to 
understand how said approximation affects these identities. 

To proceed, let us first establish that energy conservation is actually attached to  
the structure of $\Sigma_{FC}$ rather than to its explicit form, as this will prove essential 
to obtain a general principle applicable to both the approximated and non-approximated equations of motions. In the following, we therefore only assume that:
1) $\Sigma_{FC}$ is some general function of $(\eta,S,p,\Phi)$; 2) that the density entering the continuity equation (\ref{continuity}) relates to the pressure derivative of $\Sigma_{FC}$ via 
\begin{equation}
    \rho \frac{\partial \Sigma_{FC}}{\partial p}=1.  \label{rho_Sigma}
\end{equation}
Thus, taking the scalar product of (\ref{crocco_nse}) with ${\bf v}$, using the facts that by definition ${\bf v}\cdot \nabla C = DC/Dt -\partial C/\partial t = \dot{C} - \partial C/\partial t$ for any scalar field $C({\bf x},t)$ and that $\partial \Phi/\partial t=0$, yields
\begin{equation}
\begin{split} 
     \frac{D}{Dt} 
     \left ( \frac{{\bf v}^2}{2} + \Sigma_{FC} \right ) 
     = & {\bf F}\cdot {\bf v} 
     + \frac{\partial \Sigma_{FC}}{\partial \eta} \dot{\eta} 
     + \frac{\partial \Sigma_{FC}}{\partial S} \dot{S} 
     + \frac{\partial \Sigma_{FC}}{\partial t}
     - \frac{\partial \Sigma_{FC}}{\partial \eta} 
     \frac{\partial \eta}{\partial t}
     - \frac{\partial \Sigma_{FC}}{\partial S} 
     \frac{\partial S}{\partial t}  \\ 
     = & {\bf F}\cdot {\bf v} +
     \frac{\partial \Sigma_{FC}}{\partial \eta} \dot{\eta}
     + \frac{\partial \Sigma_{FC}}{\partial S} \dot{S} 
     + \frac{\partial \Sigma_{FC}}{\partial p} \frac{\partial p}{\partial t} .
\end{split} 
\end{equation}
Next, multplying the result by $(\partial \Sigma_{FC}/\partial p)^{-1} = \rho$, accounting for the continuity equation (\ref{continuity}), yields
\begin{equation}
     \frac{\partial}{\partial t} 
     ( \rho E_{total} )  
     + \nabla \cdot \left [ \rho 
     \left ( \frac{{\bf v}^2}{2} + \Sigma_{FC} 
     \right ) {\bf v} \right ] 
     = \rho \left ( {\bf F} \cdot {\bf v} 
     + \frac{\partial \Sigma_{FC}}{\partial \eta} \dot{\eta} 
     + \frac{\partial \Sigma_{FC}}{\partial S} 
     \dot{S} \right ) 
     \label{total_energy_conservation}
\end{equation}
with $E_{total}= E_k + E_p$, with $E_k = {\bf v}^2/2$ and 
\begin{equation}
   E_p = \Sigma_{FC} - p \frac{\partial \Sigma_{FC}}{\partial p} .
\end{equation}
Eq. (\ref{total_energy_conservation}) makes it clear that $E_{total}$ is conserved in the absence of viscous $({\bf F}=0)$ and diabatic ($\dot{\eta}=\dot{S}=0)$ effects. 
For total energy conservation to hold in the general case, it is necessary to make the extra assumption that the right-hand side of  (\ref{total_energy_conservation}) may be written as the divergence of some flux ${\bf J}_E$ including viscous and diffusive effects, viz., 
\begin{equation}
      \rho \left ( {\bf F} \cdot {\bf v} 
      + \frac{\partial \Sigma_{FC}}{\partial \eta} \dot{\eta} 
      + \frac{\partial \Sigma_{FC}}{\partial S} \dot{S} \right ) 
      = - \nabla \cdot ( \rho {\bf J}_E )  .
      \label{flux_constraint} 
\end{equation}
Physically, Eq. (\ref{flux_constraint}) in turn constrains the form of 
$\dot{\eta}_{irr}$ in terms of the molecular fluxes ${\bf J}_{\eta}$ and ${\bf J}_s$ as well as with the gradients of $\nabla T$ and $\nabla \mu$ to ensure consistency with the second law of thermodynamics, e.g., see \citet{Pauluis2008}, \citet{Tailleux2010} and \citet{Woods1975}. 
Note that (\ref{total_energy_conservation}) 
may alternatively be written in the form
\begin{equation}
     \frac{\partial (\rho B_{FC})}{\partial t} 
 + \nabla \cdot ( \rho B_{FC} {\bf v} ) 
 = \rho \frac{DB_{FC}}{Dt} 
 = -\nabla \cdot ( \rho {\bf J}_E) + \frac{\partial p}{\partial t}
 \label{Bernoulli_form}
\end{equation}
where
\begin{equation}
      B_{FC} = \frac{{\bf v}^2}{2} + \Sigma_{FC} 
\end{equation}
is the standard Bernoulli function attached to $\Sigma_{FC}$. 
Eq. (\ref{Bernoulli_form}) is the appropriate form of the energy conservation principle for establishing the Bernoulli theorem, namely that $B_{FC}$ is conserved along adiabatic, isohaline, inviscid, steady fluid parcel trajectories. 

\section{Pseudo-\textcolor{black}{Incompressible (PI)} approximation}
\label{pseudo_incompressible} 

To summarise, we established that under the conditions stated above, the equations of motion formulated in terms of the static energy $\Sigma_{FC}(\eta,S,p,\Phi)$ as per (\ref{normal_nse}) or (\ref{crocco_nse}), together with the continuity equation (\ref{continuity}), the entropy and salinity budgets (\ref{entropy_budget}-\ref{salinity_budget}) and the constraint (\ref{rho_Sigma}) conserve the total energy $E_{total} = E_k + E_p$, with the potential energy being given by
\begin{equation}
      E_p = \Sigma_{FC} 
      - p \frac{\partial \Sigma_{FC}}{\partial p} 
\end{equation}
regardless of the exact form of $\Sigma_{FC}$. As a consequence, the same result must hold for any approximation to the equations of motion obtained by any structure-preserving approximation of $\Sigma_{FC}$, thus providing a systematic procedure to derive such approximations, baptised here static energy asymptotics.
In this section, this approach is illustrated by re-deriving the pseudo-incompressible (PI) approximation \citep{durran_improving_1989, Durran2008, Dewar2016, Eldred2022} via replacing $\Sigma_{FC}$ by 
an approximation $\Sigma_{PI}$ defined in terms of the leading order terms of a Taylor series expansion around the reference pressure $p_R(\Phi)$ 
as follows: 
\begin{equation}
     \Sigma_{FC} = \underbrace{h(\eta,S,p_R(\Phi)) + \nu(\eta,S,p_R(\Phi)) 
     (p-p_R(\Phi)) + \Phi}_{\Sigma_{PI}}  + R_{PI} ,
     \label{sigma_pi_definition} 
\end{equation}
where $\delta p = p-p_R(\Phi)$ is the assumed small pressure anomaly and
$R_{PI} = O(\delta p^2)$ the residual term. Importantly, Eq. (\ref{sigma_pi_definition}) shows that $\Sigma_{PI} =\Sigma_{PI}(\eta,S,p,\Phi)$ 
retains the same functional form as $\Sigma_{FC}$ and hence that the above results must apply. Evaluating the $(\eta,S,p,\Phi)$ derivatives of $\Sigma_{PI}$ yields
\begin{equation}
        \frac{\partial \Sigma_{PI}}{\partial p} 
         = \nu(\eta,S,p_R(\Phi)) = \nu_{\star} = \frac{1}{\rho_{\star}}  ,
\end{equation}
\begin{equation}
        \frac{\partial \Sigma_{PI}}{\partial \Phi} 
        =  1  + \frac{\rho_R(\Phi) \delta p}{\rho^2_{\star}c^2_{s\star}} , 
\end{equation}
\begin{equation}
       \frac{\partial \Sigma_{PI}}{\partial \eta} 
        = T(\eta,S,p_R(\Phi)) + 
        \delta p \frac{\partial \nu_{\star}}{\partial \eta} = T_{\star} 
        + \delta p \frac{\partial \nu_{\star}}{\partial \eta} = T_{PI} ,
\end{equation}
\begin{equation}
      \frac{\partial \Sigma_{PI}}{\partial S} 
       = \mu(\eta,S,p_R(\Phi)) + 
       \delta p \frac{\partial \nu_{\star}}{\partial S}
        = \mu_{\star} + \delta p 
        \frac{\partial \nu_{\star}}{\partial S}  = \mu_{PI} ,  
\end{equation}
and show that these remain close to the derivatives of $\Sigma_{FC}$, with \textcolor{black}{relatively} minor impact on $\rho$, $T$, and $\mu$, the main effect being the actual pressure $p$ being replaced by the reference pressure $p_R(\Phi)$ and the presence of a small correction $O(\delta p)$, 
where $\rho_R(\Phi) = - dp_R/d\Phi$ denotes the reference density in hydrostatic equilibrium with $p_R$, \textcolor{black}{while $c_{s\star} = (\partial \rho/\partial p)^{-1/2}(\eta,S,p_R(\Phi))$ is the speed of sound evaluated at $p_R(\Phi)$. Generally, the star subscript refers to quantities evaluated at the reference pressure $p_R(\Phi)$ instead of the full pressure $p$. Note that the linearity of $\Sigma_{PI}$ in $p$ makes  $\nu_{\star} = \partial \Sigma_{PI}/\partial p$ independent of $p$, which filters sound waves out due to the resulting decoupling of density and pressure}. As a result, the above approximation affects the Crocco theorem, momentum balance equation, and continuity equation as follows:  
\begin{equation}
      \nabla \Sigma_{PI} - T_{PI} \nabla \eta - \mu_{PI} \nabla S
       = \frac{1}{\rho_{\star}} \nabla p 
       + \frac{\partial \Sigma_{PI}}{\partial \Phi} \nabla \Phi ,
\end{equation}
\begin{equation}
     \frac{D{\bf v}}{Dt} + 2 {\bf \Omega} \times {\bf v} 
      + \frac{1}{\rho_{\star}} \nabla p + 
       \frac{\partial \Sigma_{PI}}{\partial \Phi} \nabla \Phi = 
 {\bf F} ,
\end{equation}
\begin{equation}
      \frac{\partial \rho_{\star}}{\partial t} 
      + \nabla \cdot ( \rho_{\star} {\bf v} ) =  0 ,  
\end{equation}
noting that, per \eqref{rho_Sigma} applied to $\Sigma_{PI}$ instead of $\Sigma_{FC}$, $\rho$ must also be replaced by $\rho_{\star}$ in the \textcolor{black}{tracer} equations for specific entropy and salinity (\ref{entropy_budget}) and (\ref{salinity_budget}). Since $\Sigma_{PI}$ has the same functional structure as $\Sigma_{FC}$, it follows that the results derived in the previous section apply and hence that the associated approximations of motion conserve the total energy $E_{total} = E_k + E_{p,PI}$, with 
\begin{equation}
     E_{p,PI} = \Sigma_{PI} - \frac{p}{\rho_{\star}} 
      = h(\eta,S,p_R) - \frac{p_R}{\rho_{\star}} + \Phi 
       = e(\eta,S,p_R) + \Phi
       \label{energy_PI}
\end{equation}
provided that the diabatic terms $\dot{\eta}$ and $\dot{S}$ are constrained to satisfy
\begin{equation}
     \rho_{\star} [ {\bf F}\cdot {\bf v} + T_{PI} \dot{\eta} 
      + \mu_{PI} \dot{S} ] = -\nabla \cdot ( \rho_{\star} {\bf J}_E ) .
\end{equation}
In that case, (\ref{energy_PI}) shows that the approximation to the potential energy $E_{p,PI}$ is close to its exact counterpart, and hence that the energetics of the PI approximation is traceable to that of the fully compressible equations. Eq. (\ref{energy_PI}) shows that for the PI approximation to be consistent with the second law, it is sufficient to replace $T$, $\mu$, and $\rho$ by $T_{PI}$, $\mu_{PI}$, and $\rho_{\star}$ in the expressions for the diabatic terms $\dot{\eta}$ and $\dot{S}$, as previously found by \citet{Klein2012} and \citet{Eldred2021}.  

\subsection{Remarks on the choice of $p_R(\Phi)$}

Physically, the accuracy of the PI approximation is determined by the magnitude of the residual $R_{PI}$ in (\ref{sigma_pi_definition}), which may be rewritten in the following equivalent forms:
\begin{equation}
\begin{split} 
      R_{PI} = &  h(\eta,S,p)-h(\eta,S,p_R(\Phi)) 
      - \nu(\eta,S,p_R(\Phi))  (p-p_R(\Phi)) \\
      = & \int_{p_R(\Phi)}^p [ \nu(\eta,S,\tilde{p})
      - \nu(\eta,S,p_R(\Phi)) ] \,{\rm d}\tilde{p}  
      = \int_{p_R(\Phi)}^{p} \int_{p_R(\Phi)}^{\tilde{p}} 
      \nu_p (\eta,S,\tilde{\tilde{p}}) \,{\rm d}\tilde{\tilde{p}} 
      {\rm d} \tilde{p} .
      \label{r1_residual} 
\end{split} 
\end{equation}
Because $\nu_p = -1/(\rho^2 c_s^2) <0$ where $c_s$ is the speed of sound, it follows that $R_{PI}$ is also negative. Moreover, it is also easily shown that $R_{PI}$ is quadratic in $\delta p$ at leading order, viz., 
\begin{equation}
  R_{PI} \approx - \frac{(p-p_R(\Phi))^2}{2 \rho_{\star}^2 c_{s{\star}}^2} ,
  \label{r1_approximation} 
\end{equation}
and hence that its magnitude is directly controlled by the choice of $p_R(\Phi)$ and its distance to $p$. The reader familiar with the local theory APE will may recognise that \textcolor{black}{$R_{PI}$} is approximately equal to the opposite of 
\textcolor{black}{the} 
available compressible energy (ACE) denoted by $\Pi_1$ in \citet{Tailleux2018}. As a consequence,
\begin{equation}
      \Sigma_{PI} = h(\eta,S,p_R(\Phi)) + \Phi + \frac{\delta p}{\rho_{\star}} 
      = \Sigma_{FC} - R_{PI} \ge \Sigma_{FC} ,
\end{equation}
which establishes, perhaps counter-intuitively, that $\Sigma_{PI}$ represents an {\em overestimate} of the true $\Sigma_{FC}$. 
\textcolor{black}{The construction of $\Sigma_{PI}$ requires identifying a suitable reference pressure $p_R(\Phi)$, but how best to do so in practice is rarely discussed.} The fact that $R_{PI}$ is sign definite suggests, however, that $p_R(\Phi)$ should be defined to minimise the volume integral of $R_{PI}$. In the case where the denominator $\rho_{\star}^2 c_{s\star}^2$ can be treated as approximately constant, this would amount to \textcolor{black}{defining} $p_R(\Phi)$ as the horizontal-mean pressure.  

\subsection{Degenerate energy conservation principles and uniqueness issues} 

As is well known, the energy associated with the law of energy conservation for fully compressible fluids described by the NSE is $E_{tot} = E_k + E_p$, and therefore represents the fundamental form of energy against which to assess the energy conservation principles attached to any approximation to the equations of motion. 
The main aim of this section is to demonstrate that a degenerate energy conservation principle may occasionally be obtained in the context of the approximations discussed in this paper, which 
\textcolor{black}{is important to be understood} 
and be aware of for correctly interpreting some results of the literature.

As shown below, \textcolor{black}{an} alternative energy conservation principle may be shown to exist in the PI approximation because 
\begin{equation}
      \Sigma_{PI} = h(\eta,S,p_R(\Phi)) + 
      \nu(\eta,S,p_R(\Phi)) \delta p  
      + \Phi = \hat{\Sigma}_{PI}(\eta,S,\delta p,\Phi) 
      \label{sigmai_explicit}
\end{equation} 
(from Eq. (\ref{sigma_pi_definition}))
can alternatively be interpreted as a function of $(\eta,S,\delta p,\Phi)$ (denoted with a hat). Since $\delta p = p-p_R(\Phi)$ is a function of $p$ and $\Phi$ only, the pressure-geopotential gradient force can therefore be written equivalently 
as
\begin{equation}
     \frac{\partial \Sigma_{PI}}{\partial p} \nabla p
     + \frac{\partial \Sigma_{PI}}{\partial \Phi}  \nabla \Phi
     = \frac{\partial \hat{\Sigma}_{PI}}{\partial \delta p}
     \nabla \delta p + \frac{\partial \hat{\Sigma}_{PI}}{\partial \Phi} \nabla \Phi ,
     \label{PI_momentum_identity}
\end{equation}
the $(\delta p,\Phi)$ partial derivatives of 
$\hat{\Sigma}_{PI}$ being given by
\begin{equation}
      \frac{\partial \hat{\Sigma}_{PI}}{\partial \delta p} = 
      \frac{\partial \Sigma_{PI}}{\partial p} = 
      \frac{1}{\rho_{\star}}, 
\end{equation}
\begin{equation} 
      \frac{\partial \hat{\Sigma}_{PI}}{\partial \Phi} =
      1 - \frac{\rho_R(\Phi)}{\rho_{\star}} 
       + \frac{\delta p \rho_R(\Phi)}{\rho_{\star}^2 c_{s\star}^2} ,
\end{equation}
while in contrast its thermodynamic derivatives remain unaffected
\begin{equation}
     \frac{\partial \hat{\Sigma}_{PI}}{\partial \eta} = 
     \frac{\partial \Sigma_{PI}}{\partial \eta} = T_{PI} , 
     \qquad \frac{\partial \hat{\Sigma}_{PI}}{\partial S} =
     \frac{\partial \Sigma_{PI}}{\partial S} = \mu_{PI} . 
\end{equation}
From (\ref{PI_momentum_identity}), it follows that the PI momentum balance equations may alternatively be written in the form
\begin{equation}
     \frac{D{\bf v}}{Dt} + 2 {\bf \Omega}\times {\bf v} 
      + \frac{1}{\rho_{\star}} \nabla \delta p 
        + \frac{\partial \hat{\Sigma}_{PI}}{\partial \Phi} \nabla \Phi 
         = {\bf F} ,
\end{equation}
which from the same structural arguments developed previously implies the existence of a degenerate form of energy conservation for the pseudo total energy $E_{total} = E_k + E_{p,pseudo}$,
with 
\begin{equation}
        E_{p,pseudo} = \hat{\Sigma}_{PI} - 
        \delta p \frac{\partial \hat{\Sigma}}{\delta p} 
         = \hat{\Sigma}_{PI} - 
        \frac{\delta p}{\rho_{\star}} 
         = h(\eta,S,p_R(\Phi)) + \Phi .
         \label{pseudo_pe_first_encounter}
\end{equation}
However,
\begin{equation}
     \Sigma_{PI}  - p \frac{\partial \Sigma_{PI}}{\partial p}
     \ne \hat{\Sigma}_{PI} - \delta p \frac{\partial \hat{\Sigma}_{PI}}{\delta p} 
\end{equation}
so that $E_{p,pseudo}$ and $E_{p,PI}$ differ, specifically by the term \textcolor{black}{$p_R(\Phi)/\rho_{\star}$}. However, as the latter quantity may be shown to satisfy the conservation law 
\begin{equation}
       \rho_{\star} \frac{D}{Dt} \left ( \frac{p_R(\Phi)}{\rho_{\star}} \right ) 
        = \nabla \cdot ( p_R (\Phi) {\bf v} ) ,
        \label{pv_work_conservation}
\end{equation}
it follows that the conservation principle for $E_k + E_{p,pseudo}$ simply results from adding (\ref{pv_work_conservation}) to the conservation principle for the correct energy $E_k + E_{p,PI}$. This particular example highlights that achieving a flux form conservation law for an energy-like quantity is not sufficient to establish the energy conservation principle satisfied by the approximation considered: one also needs to be able to relate the energy-like quantity obtained to that of the fully-compressible energy, a point that many authors do not appear to be aware of, see Eq. (26) of \citet{Dewar2016} for instance.

\section{Anelastic Approximation (AN) revisited}
\label{generalised_anelastic} 

In the PI approximation, the density $\rho_{\star} = \rho(\eta,S,p_R(\Phi))$ \footnote{By density, we always mean the density field that obeys the continuity equation (\ref{continuity})} becomes decoupled from pressure but still responds to changes in salinity and entropy. It therefore remains time-dependent, unlike in the Boussinesq approximation. In the anelastic (AN) approximation, which we revisit in this section, the density is further approximated to be a static function of $\Phi$ only, $\rho_{\ddag}(\Phi)$. The attendant static energy $\Sigma_{AN}$ is obtained by further approximating $\Sigma_{PI}$ as follows
\begin{equation}
      \Sigma_{AN} = h(\eta,S,p_R(\Phi)) + 
      \frac{\delta p}{\rho_{\ddag}} + \Phi ,
      \label{sigma_an_definition} 
\end{equation}
and is obtained by \textcolor{black}{replacing} $\rho_{\star}=\rho(\eta,S,p_R(\Phi))$ 
by $\rho_{\ddag}(\Phi)$ 
in (\ref{sigma_pi_definition}). Per (\ref{rho_Sigma}), the continuity equation applies to $\rho_\ddag(\Phi)$, viz.:
$\nabla \cdot ( \rho_{\ddag} {\bf v} ) = 0$. Physically, this approach unifies  
(i) the standard anelastic approximation where
$\rho_{\ddag} = \rho_{R}(\Phi)$ is a general function of $\Phi$ \citep{Ogura1962, Pauluis2008, Eldred2021}  and (ii) $\rho_{\ddag} = \rho_R = {\rm constant}$, corresponding to a modernised form of Boussinesq approximation. 

The AN approximation $\Sigma_{AN}$ retains the same 
$(\eta,S,p,\Phi)$ functional dependence characterising $\Sigma_{FC}$ and $\Sigma_{PI}$. Its partial derivatives are: 
\begin{equation}
        \frac{\partial \Sigma_{AN}}{\partial p} 
         =  \frac{1}{\rho_{\ddag}}  ,
         \label{dsigman_dp}
\end{equation}
\begin{equation}
        \frac{\partial \Sigma_{AN}}{\partial \Phi} 
        = 1+ 
        \rho_R \left( \nu_\ddag - \nu_\star\right)  + \delta p \frac{\mathrm d \nu_\ddag }{\mathrm d \Phi}, 
        \label{dsigman_dz} 
\end{equation}
\begin{equation}
       \frac{\partial \Sigma_{AN}}{\partial \eta} = T_{\star},
       \label{dsigman_dn}
\end{equation}
\begin{equation}
      \frac{\partial \Sigma_{AN}}{\partial S} = \mu_{\star} ,  
      \label{dsigman_ds} 
\end{equation}
(with $\nu_\ddag=\rho_\ddag^{-1}$) and can be seen to remain close to the derivatives of $\Sigma_{PI}$, albeit with a loss of accuracy due to the absence of the small correction $O(\delta p)$ and to $\rho_{\star} \neq\rho_\ddag$.

The fact that $\Sigma_{AN}$ has the same structure as $\Sigma_{FC}$ or $\Sigma_{PI}$ means that provided that its diabatic and viscous terms satisfy the constraint
\begin{equation} 
\rho_{\ddag} [ {\bf F} \cdot {\bf v} + T_{\star} \dot{\eta} + \mu_{\star} \dot{S} ] = 
- \nabla \cdot (\rho_{\ddag} {\bf J}_E ) 
\end{equation}
analogous to (\ref{conservative_constraint_on_diabatism}), 
it must conserve the total energy $E_k + E_{p,AN}$, with the AN potential energy being given by
\begin{equation}
     E_{p,AN} = \Sigma_{AN} - p \frac{\partial \Sigma_{AN}}{\partial p} 
     = h(\eta,S,\textcolor{black}{p_R(\Phi)}) 
     - \frac{p_R}{\rho_{\ddag}} + \textcolor{black}{\Phi} 
     = E_{p,PI} + p_R (\nu_{\star} - \nu_{\ddag}) . 
     \label{AN_potential_energy}
\end{equation}

Eq. (\ref{AN_potential_energy}) shows that $E_{p,AN}$ differs only from $E_{p,PI}$ by the 
the energy error term $\Delta E_{p,spurious} = p_R (\nu_{\star} - \nu_{\ddag})$. 
This error can be minimised by defining $\nu_{\ddag}$ as the temporally and horizontally averaged value of $\nu_{\star}$ for instance.  

Eqs. (\ref{dsigman_dp}-\ref{dsigman_ds}) show that the CVT can be written in the standard form
\begin{equation}
   \nabla \Sigma_{AN} - T_{\star} \nabla \eta - \mu_{\star} \nabla \mu
   = \frac{1}{\rho_{\ddag}} \nabla p + \frac{\partial \Sigma_{AN}}{\partial \Phi} \nabla \Phi . 
   \label{AN_CVT}
\end{equation}
However, a more convenient and concise expression of the CVT can be obtained by taking advantage of the possibility of regarding $\Sigma_{AN} = \hat{\Sigma}_{AN} (\eta,S,\delta p/\rho_{\ddag},\Phi)$ as a function of $(\eta,S,\delta p/\rho_{\ddag},\Phi)$ instead, for which the $(\eta,S,\Phi)$ derivatives can be shown to be
\begin{equation}
   \frac{\partial \hat{\Sigma}_{AN}}{\partial \eta} 
   = \frac{\partial \Sigma_{AN}}{\partial \eta} = 
   T(\eta,S,\textcolor{black}{p_R(\Phi)}) = T_{\star} , 
\end{equation}
\begin{equation}
     \frac{\partial \hat{\Sigma}_{AN}}{\partial S} = 
     \frac{\partial \Sigma_{AN}}{\partial S} = 
     \mu(\eta,S,\textcolor{black}{p_R(\Phi)}) = \mu_{\star} , 
\end{equation}
\begin{equation}
       \frac{\partial \hat{\Sigma}_{AN}}{\partial \Phi} 
        = 1 - \textcolor{black}{\rho_R(\Phi) 
        \nu(\eta,S,p_R(\Phi)) =
         1 - \frac{\rho_R(\Phi)}{\rho_{\star}}} 
        = - \frac{b_{AN}}{g}  
\end{equation}
where $b_{AN}$ is recognised as the `buoyancy' defined relative to $\rho_R(\Phi)$, which differs from the traditional Boussinesq buoyancy as further discussed below. It follows that the CVT 
(\ref{AN_CVT}) may be rewritten in the form  
\begin{equation}
      \nabla \hat{\Sigma}_{AN} - T_{\star} \nabla \eta  - 
      \mu_{\star} \nabla S
      = \nabla \left ( \frac{\delta p}{\rho_{\ddag}} \right ) 
       - \frac{b_{AN}}{g} \nabla \Phi .
     \label{crocco_heuristic}
\end{equation}
Eq. (\ref{crocco_heuristic}) differs from (\ref{AN_CVT}) in that the density $\rho_{\ddag}$ is now absorbed within the anomalous pressure gradient term, while only the buoyancy $b_{AN}$ multiplies $\nabla \Phi$, thus making it possible to rewrite the standard form of momentum balance in the form
\begin{equation}
       \frac{D{\bf v}}{Dt} + 2 {\bf \Omega} \times {\bf v} 
        + \nabla \left ( \frac{\delta p}{\rho_{\ddag}} \right ) 
        - \frac{b_{AN}}{g} \nabla \Phi = {\bf F} .
        \label{standard_momentum} 
\end{equation}
The resulting diabatic anelastic system coincides with the one derived by \citet{Eldred2021}, Eq. \textcolor{black}{(4.36)}. 
A key advantage is to make the form of hydrostatic balance similar to that of the standard Boussinesq approximation even in the case where $\rho_{\ddag}$ depends on $z$, as further discussed in the next section.

\section{Modernised anelastic and Boussinesq seawater approximations} 
\label{modernised_sba} 

As established previously, the AN approximation provides an energetically and thermodynamically consistent model with 
realistic thermodynamic potentials that closely resemble their exact counterparts. In the context of 
hydrostatic ocean modelling that has been traditionally studied using legacy SBA primitive
equations models discussed in Section \ref{seawater_boussinesq_approximation}(b), 
the AN approximation takes the form (reverting to using $z$ rather than $\Phi$, which is more standard in oceanographic practice) 
\begin{equation}
      \frac{D{\bf u}}{Dt} + f {\bf k}\times {\bf u} 
       + \nabla_h \left ( \frac{\delta p}{\rho_{\ddag}} \right ) 
       = {\bf F}_h ,
       \label{AN_momentum} 
\end{equation}
\begin{equation}
      \frac{\partial}{\partial z} 
      \left ( \frac{\delta p}{\rho_{\ddag}} \right ) = b_{AN} ,
      \label{AN_hydrostatic} 
\end{equation}
\begin{equation}
       b_{AN} = - g \left ( 1 - \rho_R \nu_{\star} \right ) , 
       \label{AN_buoyancy} 
\end{equation}
\begin{equation}
      \nabla \cdot ( \rho_{\ddag} {\bf v} ) = 0  ,
      \label{AN_continuity}
\end{equation}
\begin{equation}
     \frac{DS}{Dt} = {\bf F}_S, \qquad \frac{DS}{Dt} = {\bf F}_{\theta} ,
     \label{AN_tracers} 
\end{equation}
where $\rho_{\ddag}(z)$ does
not need to coincide with $\rho_R(z) = -p_R'(z)/g$. In practice, however, it seems natural to equate
the two to avoid using multiple reference densities.
While in the standard SBA 
the reference pressure has been commonly approximated as $p_R(z) = -\rho_b g z$, 
with $\rho_b = {\rm constant}$, arguments developed in next section suggest that $p_R(z)$ and $\rho_R(z)$ could be advantageously defined in terms of the reference profiles entering the local theory of available potential energy (APE).

Mathematically, the AN model (\ref{AN_momentum}-\ref{AN_tracers}) can be verified to have essentially
the same structure as \textcolor{black}{that of} 
the legacy SBA model (\ref{SBA_momentum}-\ref{SBA_eos}), 
and hence that it can be solved algorithmically and numerically in exactly the same way regardless of whether $\rho_{\ddag}$ is taken as constant or function of $z$. 
The subtle difference between (\ref{AN_momentum}-\ref{AN_tracers}) and (\ref{SBA_momentum}-\ref{SBA_eos}) is that in (\ref{AN_buoyancy}) $b_{AN}$ represents the non-approximated buoyancy advocated by \citet{Young2012} and \citet{Tailleux2012}. Further advantages of the AN model are: 1) 
the form of potential
energy entering its energy conservation principle, viz., 
\begin{equation}
       E_{p,AN} = h(S,\theta,p_R(z)) + \Phi(z) - \frac{p_R(z)}{\rho_R(z)}  
\end{equation}
is traceable to its exact counterpart without the need to introduce ad-hoc thermodynamic potentials,
an undesirable feature of the legacy SBA models; 2) the physical basis for specifying the reference density profile $\rho_R(z)$ that it uses is much clearer than that for specifying the constant 
Boussinesq reference density $\rho_b$ used by legacy SBA models. For this reason, the above AN model
is more capable of exploiting the new capabilities offered by the Gibbs Sea Water library developed as part of TEOS-10 to its full extent. Although $\rho_{\ddag}$ could also be chosen as a constant reference Boussinesq density $\rho_b$, this simplification offers no advantage over using a depth-dependent $\rho_R(z)$ consistent with $p_R(z)$. This clearly shows, therefore, that some of the approximations made as part of the legacy Seawater Boussinesq approximation deteriorate its energetics without leading to real simplifications or computational benefits compared to the modernized version proposed here and to the anelastic system.

\section{Dynamics/thermodynamics partitioning of energy}
\label{ape_sba} 

It is now well established that energy comes in at least two different flavours, heat-like versus work-like, or available versus non-available, \textcolor{black}{or available (APE) and background (BPE)}, as justified by \citet{Lorenz1955} theory of available potential energy (APE) for instance. \textcolor{black}{Since according to thermodynamics, `work' and `heat' represent fundamentally distinct forms of energy, which in APE theory are regarded as dynamically active and inert respectively}, it follows that 
any discussion of energetics and thermodynamics consistency cannot be complete without understanding how sound-proof approximations individually affect the \textcolor{black}{active and passive forms of energy}. 

To express the above idea in the present framework, we assume that it is possible to meaningfully
decompose the static energy 
into dynamically active $(\Sigma_{\rm dyn})$ and passive $(\Sigma_{\rm heat})$ components as follows:
\begin{equation}
     \Sigma_{FC} (\eta,S,p,\Phi) = \Sigma_{\rm dyn} (\eta,S,p,\Phi) 
      + \Sigma_{\rm heat} (\eta,S) .
\end{equation}
Since $\Sigma_{\rm heat}$ is independent of $p$ and $\Phi$, it follows that only $\Sigma_{\rm dyn}$ is
dynamically relevant for predicting the pressure-geopotential gradient force entering the standard form of momentum balance equations, 
\begin{equation}
    \frac{1}{\rho} \nabla p + \nabla \Phi =
    \frac{\partial \Sigma_{\rm dyn}}{\partial p} \nabla p 
    + \frac{\partial \Sigma_{\rm dyn}}{\partial \Phi} \nabla \Phi ,
    \label{sigma_dynamical} 
\end{equation}
thus justifying the dynamically irrelevant character of $\Sigma_{\rm heat}$.
In the following, we briefly discuss the two main approaches for defining $\Sigma_{\rm dyn}$ and $\Sigma_{\rm heat}$ proposed so far, and how these are affected by the most drastic AN approximation.
Eq. (\ref{sigma_dynamical}) shows that to be physically acceptable, any construction of $\Sigma_{\rm dyn}$ should only require knowledge of $\partial \Sigma/\partial p$ and $\partial \Sigma/\partial \Phi$ (regardless of how $\Sigma$ is defined), possibly combined with knowledge of the whole stratification at some point in time (such as provided by initial conditions).

For the sake of clarity and consistency with existing literature, we switch back to using height/depth $z$ as vertical coordinate. 

\subsection{Available potential energy (APE) theory}

From a theoretical viewpoint, the most rigorous and physically-based partitioning of $\Sigma_{FC}$ is arguably the one rooted in APE theory. In this approach, $\Sigma_{\rm dyn}$ is defined as the departure of $\Sigma_{FC}$ from its equilibrium value defined as the one it would have 
in a suitably defined notional
reference state characterised by the reference pressure $p_0(z)$ and $\rho_0(z)$, that is, 
\begin{equation}
     \Sigma_{\rm heat} = h(\eta,S,p_0(z_R)) + g z_R 
     \label{sigma_heat_ape} 
\end{equation}
where $z_R = z_R(\eta,S)$ is the fluid parcel's reference position. 
\textcolor{black}{Eq. (\ref{sigma_heat_ape}) thus represents a local form of background potential energy (BPE), which in \citet{Lorenz1955} was only defined as a volume-integrated quantity}.

Because in a state of rest,
the density of a fluid parcel must match that of the background reference state, $z_R$ must be 
a solution of the so-called level of neutral buoyancy (LNB) equation
\citep{Tailleux2013b,Tailleux2018} 
\begin{equation}
    \rho(\eta,S,p_0(z_R)) = \rho_0(z_R) .
    \label{LNB_equation} 
\end{equation}
With $\Sigma_{\rm heat}$ defined as per (\ref{sigma_heat_ape}), we have
\begin{equation}
     \Sigma_{\rm dyn} = \Sigma_{FC} - \Sigma_{\rm heat} = 
     \Pi_1 + \Pi_2 + \frac{p-p_0(z)}{\rho} 
     \label{sigma_dyn} 
\end{equation}
where $\Pi_1$ and $\Pi_2$ are so-called available compressible energy (ACE) density and APE density respectively, 
\begin{equation}
    \Pi_1 = h(\eta,S,p)-h(\eta,S,p_0(z)) 
    + \frac{p_0(z)-p}{\rho} 
\end{equation}
\begin{equation}
     \Pi_2 = h(\eta,S,p_0(z)) - h(\eta,S,p_0(z_R)) + g (z-z_R) .
\end{equation}
As shown by \citet{Tailleux2018}, these may be written as
\begin{equation}
     \Pi_1 = \int_{p_0(z)}^{p} \int_{p'}^{p} \frac{\partial \nu}{\partial p}
     (\eta,S,p'')\,{\rm d}p'' {\rm d}p' 
\end{equation}
\begin{equation}
     \Pi_2 = - \int_{z_R}^z b(\eta,S,z') \,{\rm d}z' 
\end{equation} 
where $b(\eta,S,z) = - g[1 - \rho_0(z) \nu(\eta,S,p_0(z))]$ is the buoyancy defined
relative to the reference density $\rho_0(z)$. This establishes, therefore, that $\Pi_1$, $\Pi_2$, and therefore $\Sigma_{\rm dyn}$ can be constructed only from the knowledge of $\nu(\eta,S,p)$, $g$, as well as from one state of the system from which $p_0(z)$ and $\rho_0(z)$ are constructed \textcolor{black}{(for instance by using the adiabatic and isohaline re-arrangement technique discussed by \citet{Saenz2015})}. 

Let us now seek to understand how the AN approximation affects the dynamics/thermodynamics partitioning of $\Sigma_{AN}$, 
\begin{equation}
       \Sigma_{AN} = h(\eta,S,p_R(z)) + gz + \frac{\delta p}{\rho_{\ddag}} .
\end{equation}
By considering the value of $\Sigma_{AN}$ in the Lorenz background reference state, in which
$z=z_R$ and $p=p_0(z_R)$, it is easily realised that $\Sigma_{AN,heat}$ can be made
to coincide with its exact counterpart (\ref{sigma_heat_ape}) simply by choosing $p_R(z) = p_0(z)$,
thus providing a natural solution to the problem of how to define $p_R(z)$ in the PI and AN
approximations. With such a choice, the dynamical component of $\Sigma_{AN}$ becomes
\begin{equation}
     \Sigma_{AN,dyn} = 
     \Sigma_{AN} - \Sigma_{AN,heat} = \Pi_2 + \frac{p-p_0(z)}{\rho_{\ddag}} ,
     \label{sigma_an_dyn} 
\end{equation}
where the expression for $\Pi_2$ is the same as its exact counterpart. 
By comparing (\ref{sigma_an_dyn}) with its exact counterpart (\ref{sigma_dyn}), it is seen that the
AN approximation results in the disappearance of the ACE component $\Pi_1$, while also approximating
the compressible work term $(p-p_0(z))/\rho$ by $(p-p_0(z))/\rho_{\ddag}$. To avoid the proliferation
of reference density profiles, the most logical choice is to use $\rho_{\ddag} = \rho_0(z)$, which 
provides a natural solution for how to choose $\rho_{\ddag}$.

\subsection{Potential/Dynamic enthalpy partitioning}

We briefly remark that another kind of partitioning is based on defining $\Sigma_{\rm heat}$ as 
the `potential' value of $\Sigma_{FC}$ referenced to the ocean surface at $z=0$, viz., 
\begin{equation}
    \Sigma_{heat} = h(\eta,S,p_R(0)) + \Phi(0) ,  
    \label{potential_enthalpy_sigma} 
\end{equation}
\citep{deSzoeke2000a,Young2010,Nycander2010}
where $p_R(z)$ is an arbitrary hydrostatic reference pressure field defining the reference density
field via $\rho_R(z) = -p_R'(z)/g$. If $p_R(0)=p_a$ and $\Phi(0)=0$, where $p_a$ is the mean surface atmospheric pressure, $\Sigma_{\rm heat}$ then 
coincides with \citet{McDougall2003}'s potential enthalpy, the currently 
recommended variable
for defining heat content in the oceans following the adoption of TEOS-10
\citep{IOC2010,Pawlowicz2012}. With $\Sigma_{\rm heat}$ defined as per (\ref{potential_enthalpy_sigma}),
the expression for $\Sigma_{\rm dyn}$ can easily be verified to be
\begin{equation}
     \Sigma_{\rm dyn} = \Pi_1 + h^{\ddag}(\eta,S,z) + \frac{p-p_R(z)}{\rho}  
     \label{sigma_dyn_enthalpy} 
\end{equation}
with 
\textcolor{black}{$\Pi_1 = h(\eta,S,p) - h(\eta,S,p_R(z)) + (p_R(z)-p)/\rho \ge 0$} 
the ACE defined in terms of $p_R(z)$ rather than $p_0(z)$, and $h^{\ddag}$ the so-called dynamic enthalpy, defined by
\begin{equation}
     h^{\ddag} = h(\eta,S,p_R(z)) - h(\eta,S,p_R(0)) + \Phi(z) - \Phi(0)
     = -\int_0^z b(\eta,S,z')\,{\rm d}z' 
\end{equation}
with $b(\eta,S,z) = - g[1-\rho_R(z) \nu(\eta,S,p_R(z))]$ the buoyancy defined relative to the reference density $\rho_R(z)$. It is easily seen that (\ref{sigma_dyn_enthalpy}) and
(\ref{potential_enthalpy_sigma}) only differ from their APE-based counterparts by the use of a constant reference level located at the surface in place of the parcels' resting position, as well as in the choice of reference density/pressure profiles. Similarly as for the APE-based
BPE, the AN approximation does not affect the thermodynamic component of $\Sigma_{AN}$, so 
that $\Sigma_{AN,heat} = \Sigma_{\rm heat}$. The AN approximation only affects the dynamical
component by getting rid of $\Pi_1$ in (\ref{sigma_dyn_enthalpy}), so that 
\begin{equation}
       \Sigma_{AN,dyn} = h^{\ddag} + \frac{\delta p}{\rho_{\ddag}} .
\end{equation} 
The potential/dynamic enthalpy based partitioning of $\Sigma_{AN}$, however, lacks any of
the advantageous attributes and clear physical interpretation of the BPE/APE based partitioning; its usefulness is therefore questionable.

\section{Summary and discussion} 
\label{summary_and_discussion} 

The main result of this paper is that it is possible to write the equations of motion for a general compressible two-constituent stratified fluid, as well as for a large class of thermodynamically and energetically consistent sound-proof approximations of such equations,
under the generic form
\begin{equation}
    \Sigma = \Sigma(\eta,S,p,\Phi) ,
\end{equation}
\begin{equation}
    B = \frac{\partial \Sigma}{\partial \Phi},
    \qquad \rho^{-1} = \nu = \frac{\partial \Sigma}{\partial p},
    \qquad T = \frac{\partial \Sigma}{\partial \eta} , 
    \qquad \mu = \frac{\partial \Sigma}{\partial S} ,
\end{equation}
\begin{equation}
     \frac{D{\bf v}}{Dt} + 2 \boldsymbol{\Omega} \times {\bf v} 
     = - \nu \nabla p 
     + B \nabla \Phi  + {\bf F} ,
\end{equation}
\begin{equation}
     \nabla \cdot {\bf v} = \frac{D}{Dt} \ln{\nu} ,
\end{equation}
\begin{equation}
      \frac{D\eta}{Dt} = - \nu
       \nabla \cdot \left ( \rho
      {\bf J}_{\eta} \right )  + \dot{\eta}_{irr} = \dot{\eta} ,
\end{equation}
\begin{equation}
      \frac{DS}{Dt} = - \nu
      \nabla \cdot \left ( \rho 
       {\bf J}_s \right )  = \dot{S} ,
\end{equation}
\begin{equation}
     {\bf F} \cdot {\bf v} + 
     T \dot{\eta} 
     + \mu \dot{S} =
     - \nu \nabla \cdot \left ( \rho {\bf J}_E  \right) ,
\end{equation}
\begin{equation}
     T {\bf J}_{\eta} = - L_{\eta \eta} \nabla T 
     - L_{\eta s} \nabla \mu ,
\end{equation}
\begin{equation}
     T {\bf J}_s = - L_{\eta s} \nabla T 
     - L_{ss} \nabla \mu .
\end{equation}
As shown in this paper, such equations conserve total mass defined as the volume integral of $\left ( \frac{\partial \Sigma}{\partial p} \right )^{-1}$, total salinity defined as the volume integral of $\left ( \frac{\partial \Sigma}{\partial p} \right )^{-1} S$ and  total energy defined as the volume integral of $\left ( \frac{\partial \Sigma}{\partial p} \right )^{-1} E_{tot}$, where $E_{tot} = {\bf v}^2/2 + \Sigma - p \partial \Sigma/\partial p$.  Furthermore, they are consistent with the second law of thermodynamics. This result holds because the sound-proof approximations considered in this paper only alter the form of $\Sigma$ but not its dependence on the core $(\eta,S,p,\Phi)$ variables. This result is important, because it demonstrates that it is possible to formulate sound-proof approximations in a way that 
\textcolor{black}{makes} their energetics and thermodynamics parallel that of the fully compressible Navier-Stokes equations, whose feasibility had remained unclear until now. For instance, \citet{Huang2001} argued that the Boussinesq equations are fundamentally affected by significant energy errors, but the present results establish that this can easily be avoided by adopting the above formalism and the proposed modernization of the Boussinesq system. 

This work underscores the fundamental importance of static energy for elucidating the energetics and thermodynamics of sound-proof approximations to the equations of motion. Physically, this is because
the static energy encapsulates all fundamental information about the potential energy and pressure-geopotential gradient force that it generates, so that energetically and thermodynamically consistent approximations can be simply constructed by approximating the static energy to various orders of accuracy while respecting a few simple rules. From a practical standpoint, this approach naturally gives rise to two well-known important types of consistent approximations: pseudo-incompressible (PI) and (modernised) anelastic (AN), but which are here re-derived more transparently and succinctly while also clarifying some aspects that are not generally discussed, such as that pertaining to the choice of reference states. 

The proposed approximation procedure is similar in many respects to Halmilton's Principle Asymptotics (HPA) advocated by \citet{holm_euler-poincare_2002} and leveraged by many authors, e.g. \citet{Vasil2013, tort_usual_2014, tort_dynamically_2014, dubos_semihydrostatic_2014, dubos_covariant_2018}. \citet{holm_euler-poincare_2002} argue that approximations should be developed directly at the level 
\textcolor{black}{of} the Lagrangian from which the adiabatic equations of motion derive via Hamilton's least action principle, because this guarantees the existence of conservation laws whenever they should hold, i.e. when the Lagrangian presents the required symmetries. For the adiabatic part of the equations of motions, our method is thus a special case of HPA, limited to approximations of the Lagrangian that modify only static energy, thus motivating the name Static Energy Asymptotics (SEA) used in this paper to refer to our method. While more restrictive than the full HPA, SEA is also more comprehensive in that it deals consistently with diabatic processes. SEA is thus a stripped-down version of HPA applied to variational principles including diabatic processes, circumventing the advanced mathematical machinery used by \citet{Eldred2021}.

In the context of numerical ocean modelling, the PI and AN approximations are both superior to the standard SBA; their potential energy and thermodynamic potentials are nearly identical to their exact counterparts so that their energetics is more easily comparable to that of \textcolor{black}{the fully compressible equations} and do not require the introduction of artificial and ad-hoc thermodynamic potentials. The PI and AN approximations can therefore leverage the full power the Gibbs Sea Water (GSW) library developed as part of the new TEOS-10 equation of state \citep{IOC2010,Pawlowicz2012}. Both approximations only affect the local APE density by removing its available compressible part $\Pi_1$ but 
\textcolor{black}{leave}
the local BPE unaffected. In contrast, how to define the local BPE for the SBA is much less clear, \textcolor{black}{given} that closing its energetics requires the construction of ad-hoc potentials. 
In relation to previous work, our seawater PI approximation (which is essentially a special case of the ``atmospheric'' PI approximation with seawater thermodynamics) provides a physically more transparent and more systematic foundation for the PI models that \citet{Dewar2016} proposed to improve on existing SBA-based OGCMs. 
As to our seawater AN approximation, it unifies the modernized-Boussinesq and anelastic approximations into a single framework, some of its elements being found in previous works by \citet{Pauluis2008,Young2010,Tailleux2012}. Importantly, it shows that to construct a seawater anelastic approximation, linearisation of the equation of state is not necessary, as previously established by \citet{Pauluis2008} 
but in contrast to what was assumed in \citet{Ingersoll2005}. From a practical viewpoint, implementations of the PI models would require non-trivial alterations to existing OGCM codes and remains to be attempted. In contrast, the seawater generalised AN approximation has essentially the same mathematical structure as that of the standard SBA, so that 
\textcolor{black}{its} 
practical implementation should be straightforward. As doing so would significantly improve the energetics and thermodynamics of existing OGCMs at little to no additional cost, we recommend that efforts should be devoted in the future to adopt it. A natural choice of reference pressure and density profiles would be the analytical profiles entering the formulation of the analytical form of thermodynamic neutral density proposed in \citet{Tailleux2021}.

We believe that our findings can demystify a traditionally intricate and obscure aspect of atmospheric and ocean modelling. The issue of whether an approximation to the equations of motion is energetically and thermodynamically consistent has often been perceived as arcane, lacking systematic and general rules. We hope that our work will help atmospheric and ocean modellers realise that this issue is simpler than previously assumed, and that our findings will contribute to enhancing the physical realism of atmospheric and oceanic models, thus addressing the concerns raised by \citet{Lauritzen2022}.

\paragraph{Acknowledgements} 
This work was supported by an International Exchange award from
the Royal Society number IES$\backslash$R3$\backslash$223161 .
This work is part of the AWACA project
that has received funding from the European Research Council (ERC) under the European Union’s Horizon 2020 research and innovation
programme (Grant agreement No. 951596). \textcolor{black}{The authors thank four anonymous reviewers for their useful and thorough remarks, which greatly helped improve the final manuscript.}

\appendix







\end{document}